\documentclass[preprint2]{aastex631}
\begin{document}
\title{Potential Vibrational Modes Tied to Diffuse Interstellar Bands}

\author[0000-0001-8803-3840]{Daniel Majaess}
\affiliation{Department of Chemistry and Physics, Mount Saint Vincent University, Halifax, Nova Scotia, B3M2J6 Canada.}
\email{Daniel.Majaess@msvu.ca}

\author{Halis Seuret}
\affiliation{Centro de Investigaciones Químicas, IICBA, Universidad Autónoma del Estado de Morelos, Cuernavaca, 62209, Morelos, Mexico.}
\affiliation{Department of Chemistry and Physics, Mount Saint Vincent University, Halifax, Nova Scotia, B3M2J6 Canada.}

\author[0000-0003-3469-8980]{Tina A. Harriott}
\affiliation{Department of Mathematics and Statistics, Mount Saint Vincent University, Halifax, Nova Scotia, B3M2J6 Canada.}
\affiliation{Department of Chemistry and Physics, Mount Saint Vincent University, Halifax, Nova Scotia, B3M2J6 Canada.}

\author[0000-0002-8746-9076]{Cercis Morera-Boado}
\affiliation{IXM-Cátedra Conahcyt-Centro de Investigaciones Químicas, IICBA, Universidad Autónoma del Estado de Morelos, Cuernavaca, 62209, Morelos, Mexico.}

\author{Ailish Sullivan}
\affiliation{Department of Chemistry and Physics, Mount Saint Vincent University, Halifax, Nova Scotia, B3M2J6 Canada.}

\author[0000-0001-6662-3428]{Lou Massa}
\affiliation{Hunter College \& the PhD Program of the Graduate Center, City University of New York, New York, USA.}

\author[0000-0001-8397-5353]{Ch\'erif F. Matta}
\affiliation{Department of Chemistry and Physics, Mount Saint Vincent University, Halifax, Nova Scotia, B3M2J6 Canada.}
\affiliation{Department of Chemistry, Saint Mary's University, Halifax, Nova Scotia, B3H3C3 Canada.}
\affiliation{D\'epartement de Chimie, Universit\'e Laval, Qu\'ebec, G1V0A6 Canada.}
\affiliation{Department of Chemistry, Dalhousie University, Halifax, Nova Scotia, B3H4J3 Canada.}

\begin{abstract}
 Potential vibrational modes associated with diffuse interstellar bands (DIBs) could be discerned by examining energy differences between correlated DIBs. Consequently, $\approx 10^3$ higher correlated DIB pairs ($r-\sigma_r \ge 0.8$, $\ge 12$ sightlines) were extracted from the Apache Point Observatory DIB catalog, and their energy spacings computed. In this first macro exploratory step, a histogram possibly reveals chemical bond signatures of C$\equiv$C, C$\equiv$N, S$-$H, C$-$O, C$=$O, Si$-$H, N$-$H, C$-$H (aliphatic), C$\mathbf{^{\underline{...}}}$C (in-ring), and aromatics (C$-$H stretch, C$\mathbf{^{\underline{...}}}$C in-ring, oop C$-$H bending, and overtones). Continued research is required to (in)validate the histogram approach, mitigate noise, scrutinize maxima, break degeneracies, and converge upon an optimal framework.
 \end{abstract}

\section{Introduction}

\citet{heg22} observed that absorption lines at 5780 and 5797 \AA ~were superposed upon the spectra of binary stars, and lacked the requisite oscillatory Doppler shifting. The source(s) of these lines lie mainly within interstellar clouds along the sightline \citep[see also][regarding interstellar calcium]{har04}. A century later several hundred diffuse interstellar bands (DIBs) are known \citep[e.g.,][]{bon12,Fan2019}. PAHs remain a leading hypothesis as a principal carrier \citep[e.g.,][]{bon20}, and for several DIBs C$_{60}^{+}$ is debated \citep[e.g.,][]{cam15,gal17,gal21,sch21,nie22,maj25}. Indeed, heterofullerenes and (endo/exo)hedral inclusions are likewise being explored as DIB carriers \citep[e.g.,][]{kr87,om16}.  

\begin{figure*}
\begin{center}
\includegraphics[width=0.95\linewidth]{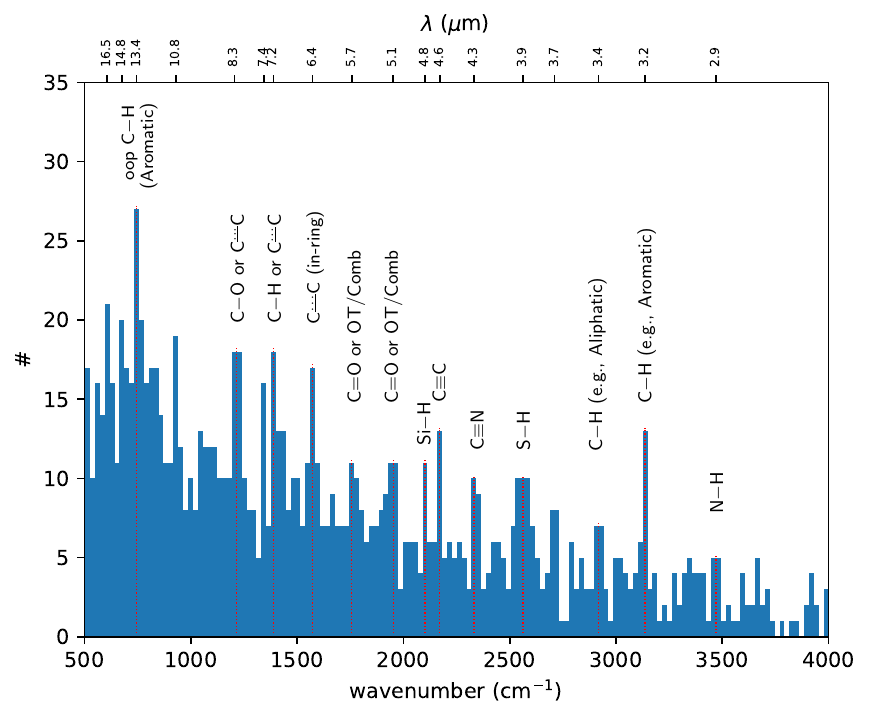}
\caption{Energy differences between higher correlated DIB pairs ($r-\sigma_r \ge 0.8$,  $EW/\sigma_{EW} \ge 5$) may feature maxima that reveal the underlying chemical bonds (tentative candidates are suggested). Degeneracies exist owing to broadening and overlapping wavenumbers. Independent investigations are needed to evaluate the histogram approach and identify spurious maxima. Data were extracted from the APO catalog of DIBs.}
\label{fig1}
\end{center}
\end{figure*}

Here, the objective is to explore whether vibrational transitions may be identified by delineating energy differences between correlated DIB pairs \citep[e.g.,][]{jd93,mo99,bon20}.  For example, \citet{jd93} suggested the energy separation between DIBs 5797 and 6269 {\AA} could be indicative of a PAH C$=$C vibration (7.7 $\mu m$).  \citet{mo99} underscored that the gap between the correlated 6196 and 6614 {\AA} DIBs is tied to an aromatic vibration (9.8 $\mu m$). \citet[][their Table 4]{bon20} relays that the energy offset between DIBs 5545 and 6614 {\AA} may be linked to a PAH or aliphatic C$-$H vibration (3.3 $\mu m$).  DIBs associated with a given molecule may represent a vibronic progression \citep[e.g.,][and discussion therein]{mcc10}.

%%%%%%%%%%%%%%%%%%%%%%%%%%%%%%%%%%%%%%%%%%%%%%%%%%%%%%%%%%%%%%%%%%%%%%%%%%%%%%%%%%%%%%
\section{Analysis}
The \citet{Fan2019} APO catalog was examined, and the analysis was subsequently limited to DIB pairs exhibiting higher Pearson correlated equivalent widths ($r-\sigma_r \ge 0.8$, $EW/\sigma_{EW} \ge 5$), possessing $\ge 12$ sightlines, and whose energy difference falls within $500-4000$ cm$^{-1}$. The Pearson correlation, equivalent width, and their uncertainties are described by $r$, $\sigma_r$, $EW$, and $\sigma_{EW}$. The sightline to VI Cyg 12 was excluded owing to its circumstellar shell and color-excess beyond the field \citep[e.g.,][]{Maryeva,hel24}. 

The final sample hosts $\simeq 10^3$ DIB pairs.  Wavenumbers linked to the energy spacing between DIB pairs were compiled into a histogram (23 cm$^{-1}$ bin width).  Vibrations were identified by relying on \citet[][]{col90}, the ChemCompute+GAMESS quantum chemistry framework \citep{per14,bar20}, and the NASA Ames PAH IR spectroscopic database \citep{boe14,ba18,mat20}. Tentatively, the peaks in Fig.~\ref{fig1} can be assigned to various chemical bonds (e.g., C$\equiv$C, C$\equiv$N, S$-$H, C$-$O, C$=$O, Si$-$H).  For example, the potential aromatic out of plane (oop) bending C$-$H vibration may represent the line near 745 cm$^{-1}$, which is the most prominent maximum,\footnote{Linearly binned wavelength (rather than wavenumber) would reveal a maximum toward small $\lambda$.} with an underestimated uncertainty (formal) being half the bin width (i.e., $745\pm12$ cm$^{-1}$). Peaks in its vicinity could represent differing aromatic substitution patterns. The prominence of $\simeq 745$ cm$^{-1}$ (13.4 $\mu m$) in concert with $\simeq697$ cm$^{-1}$ (14.8 $\mu m$) may be indicative of mono-substitution. The feature near 606 cm$^{-1}$ (16.5 $\mu m$) was identified by \citet[][]{mo00} as linked to PAHs \citep[see also][and their DIB family]{bon20}.  Aromatics are likewise relayed by the in-ring C$\mathbf{^{\underline{...}}}$C line perhaps appearing near 1573 cm$^{-1}$, and C$-$H line beyond $\simeq 3000$ cm$^{-1}$, while shortward of the latter are aliphatic C$-$H. Furthermore, overdensities near 5.25 (FWHM$\approx 0.12$ $\mu m$) and 5.7 $\mu m$ (FWHM$\approx 0.17$ $\mu m$) can be conducive to PAH emission from overtones, combinations, etc. \citep[][and references therein]{bo09a}. The two longer wavelength C$\mathbf{^{\underline{...}}}$C may be tied to fullerenes, and a degeneracy could likewise extend to the putative 10.8 $\mu m$ and oop C$-$H features. The diversity of vibrational transitions reaffirms prior analyses indicating numerous molecules give rise to DIBs \citep[e.g., on the basis of correlated equivalent widths, common correlations relative to reddening, and spectral line morphology,][]{cam97,fra21,esmith2022,eb24}.

\begin{figure*}
\begin{center}
\includegraphics[width=0.95\linewidth]{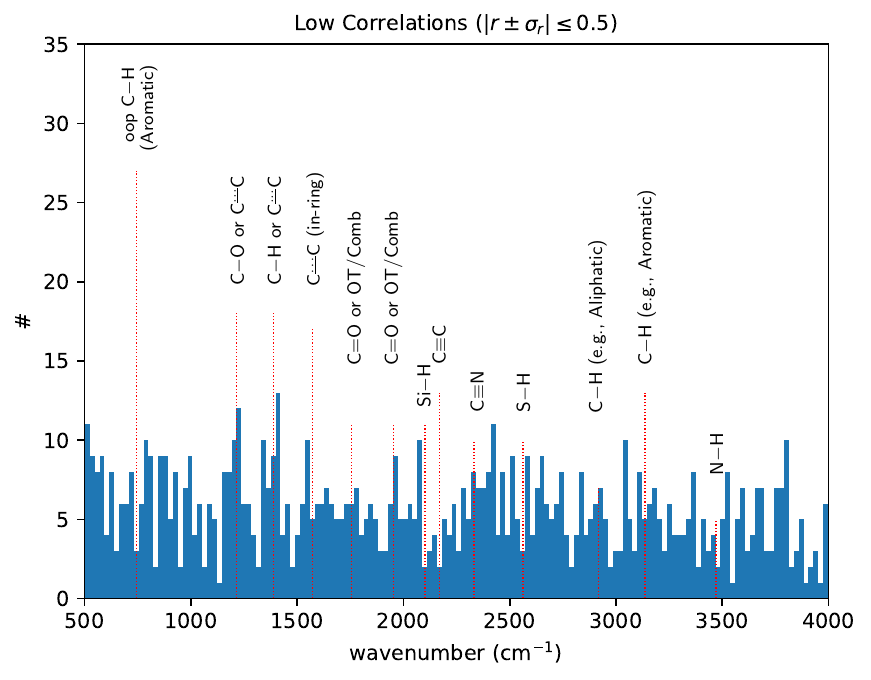}
\caption{Energy differences between low correlation DIB pairs ($|r\pm\sigma_r| \leq 0.5$).  Relative to the high correlation analysis (Fig.~\ref{fig1}), the dominant line and substructure at smaller wavenumbers are comparatively absent. Expectedly, a lower significance is apparent for a subset of vibrational modes that possibly remain.}
\label{fig2}
\end{center}
\end{figure*}

Crucially, artifacts may exist owing to noise (e.g., N$-$H), and a balance was sought where sufficient statistics were achieved in concert with a reasonable selection of the correlation threshold, sightline number, and binning. Consequently, a histogram for DIB pairs displaying low correlations was constructed (i.e., $|r\pm\sigma_r| \leq 0.5$, Fig.~\ref{fig2}) as one possible means of assessing the veracity of the maxima.  The maxima were expectedly sensitive to the criteria selected (e.g., $r-\sigma_r \ge 0.8$).  The dominant $\simeq745$ cm$^{-1}$ line that characterized higher correlated DIB pairs (Fig.~\ref{fig1}) vanishes, and the underlying substructure at smaller wavenumbers is likewise absent. A subset of vibrational modes potentially remain with less significance owing to the lower correlation criterion, with only one exceeding 3$\sigma$. The red dotted lines in Fig.~\ref{fig2} stem from the bin centers of Fig.~\ref{fig1}.  Sample sizes for Figs.~\ref{fig1} and \ref{fig2} are 1143 and 854 DIB pairs, accordingly.

Yet ultimately, the preliminary vibrations designated in Fig.~\ref{fig1} require further benchmarking and independent vetting.  Adjustments shall likewise proceed as a consensus is achieved over time, since vibrational modes can overlap, their wavelengths can shift owing to other constituents within the molecule, and broadening and degeneracies occur \citep[e.g.,][]{tru23}. 

\section{Conclusions}
In this brief exploratory note, DIB energy differences (e.g., Fig.~\ref{fig1}) may unveil the building blocks inherent to the broader host molecules.  For example, aromatics (e.g., hydrocarbons and potentially heterocycles) and fullerenes could represent a subset of DIB carriers (Fig.~\ref{fig1}), as noted previously by others. Subsequent key steps moving forward include continuing to isolate DIB families (i.e., same carrier) on a multi-dimensional basis of equivalent widths, optical and near-infrared reddening, line profiles, etc.~\citep[e.g.,][]{eb24}. Such ongoing research is required to mitigate the noise in Fig.~\ref{fig1}, which partly arises from correlated DIB pairs linked to separate carriers whose abundances are commensurate. A critical aspect is to correctly unveil the DIB tied to the origin band, which may represent the transition to the ground vibration of the first excited electronic state.\footnote{Slight offsets between observed vibrational wavenumbers implied by DIB pairs relative to those in compilations are expected if the latter are linked to the ground electronic state.} Concurrently, the APO catalog can be expanded by extracting additional EWs from high-quality GOSSS and X-shooter spectra \citep[][]{mai13,ver22}, while simultaneously characterizing the number and properties of dust clouds along the sightline by utilizing new Gaia DR3 parallax and $\lambda \simeq 330-1050$ nm spectroscopic observations \citep{Gaia,hel24}.  The latter may provide the desirable rationale behind outliers amongst Pearson correlation determinations \citep[e.g., circumstellar shell for VI Cyg 12,][their Fig.~1]{hel24}. Moreover, viewing a DIB through multiple clouds along the sightline can be preferable when establishing broad correlations, thereby mitigating anomalies endemic to any one cloud.  

Future work likewise includes awaiting temporally costly extensive vibrational coupled cluster calculations for an expansive set of neutral and cation species, and undertaking analyses of linearly binned wavelength histograms and unidentified infrared emission lines (UIEs).\footnote{\citet{kw22} favors mixed aromatic/aliphatic organic nanoparticles (MAONs) for UIEs rather than canonical PAHs.} DIBs and UIEs should share a subsample of molecules,\footnote{e.g., \citet{bon20}, and for C$_{60}^{+}$ see \citet[][DIBs]{fe94} and \citet[][UIEs]{sad22}.} however, differences are expected (e.g., $\lambda$ linked to neutral versus ion species, intensity shifts, separate molecules) owing to disparate ambient temperatures, densities, neutral and ion population ratios, radiation field, etc.~\citep[broader discussions in \citealt{pe02} and][and references therein]{bon20}.

\begin{acknowledgments}
\textbf{Acknowledgments}:~this research relied on initiatives such as the APO Catalog of DIBs, CDS, NASA ADS, arXiv, NASA Ames PAH IR spectroscopic database, ChemCompute+GAMESS. 
\end{acknowledgments}

\bibliography{article}{}
\bibliographystyle{aasjournal}
\end{document}